\documentclass[letterpaper, reqno]{amsart} 
\usepackage{epsfig} 
\usepackage{oldlfont}
\newcommand{\vev}[1]{\langle #1 \rangle} 
\newcommand{\deriv}[2]{\frac{\partial #1}{\partial #2}}
\begin{document} 
\title{On the Universality of the Entropy-Area Relation} 
\author{Seth A. Major and Kevin L. Setter} 
\date{August 2001} 
\address{Major: Department of Physics\\ 
Hamilton College\\ 
Clinton NY 13323 USA\\ 
Setter: Department of Physics and Astronomy\\ 
Swarthmore College\\ 
Swarthmore PA 19081 USA} 
\email{Major: smajor@hamilton.edu \\ Setter: ksetter1@swarthmore.edu} 

\begin{abstract} 
We present an argument that, for a large class of possible dynamics, a
canonical quantization of gravity will satisfy the Bekenstein-Hawking
entropy-area relation.  This result holds for temperatures low
compared to the Planck temperature and for boundaries with areas large
compared to Planck area.  We also relate our description, in terms of a
grand canonical ensemble, to previous geometric entropy calculations
using area ensembles.
\end{abstract} 
\maketitle

\section{Introduction}

When formulating a new physical theory, especially one as speculative
as quantum gravity, it is essential to have limiting behavior to
test possible theories.  It is also useful to understand how
tight this limiting behavior constrains the theory.  For quantum
gravity, the semiclassical Bekenstein-Hawking relation provides one
such limit.  This paper is an exploration of the nature of the
constraint due to the entropy-area relation.  We find, in fact, that
the restriction is extremely mild.

To make this argument, we use a model motivated by the formulation of
geometry in loop quantum gravity.  In this formulation, the states of
the gravitational field are given by intersecting one-dimensional
excitations of geometry called spin networks.  States of a surface,
such as an horizon, are determined by the quantum states at the
intersections of the spin network and the surface.  These
intersections or ``geometric particles'' in turn determine the
resulting statistical mechanics.

The current status of the theoretical development of loop quantum
gravity (see Ref.  \cite{rev} for a review) is that while the
kinematics is relatively established, the dynamics remains
controversial.  One intriguing possibility is that the study a four
dimensional formulations called spin foams may resolve the form of the
Hamiltonian constraint.  Another possibility is that the correct
formulation of the Hamiltonian constraint could be settled with
acceptable limiting behavior such as the Bekenstein-Hawking relation. 
The present work suggest that the entropy-area proportionality is
largely insensitive to the microdynamics of the boundary theory. 
Instead, the relation is seen to be a consequence of the scaling
property of a class of very simple statistical systems.

A number of previous papers have studied the Bekenstein-Hawking
relation in loop quantum gravity.  Early work
\cite{carlobh,kirill1,kirill2} concentrated on computing the
geometrical entropy associated with fixed total area.  Later work
\cite{abck,abk} provided a more careful analysis of the boundary
conditions and the resulting intrinsic geometry on the horizon.  In
all these cases, the proportionality of entropy and area was
recovered.  The Bekenstein-Hawking relation could then be used to fix
the undetermined parameter in the theory.  In addition, it was found
that the black hole entropy could be accounted for by the quantum
states of the horizon geometry \cite{abk}.  More recent work by the
present authors \cite{gravstatmech} uses a definition of the
quasilocal energy to explore the statistical mechanics of a region of
space bounded by a surface.  A precise form of the Hamiltonian on
the boundary was used in that derivation.  Unlike that work, in this
paper we find that a wide class of possible Hamiltonians also recover the
Bekenstein-Hawking relation.

Many properties of extremal and near extremal black holes
have been derived in string theory.  (See Ref.  \cite{apeet} for a
review.)  It would be interesting to explore the universality of the
entropy-area relation in that context as well.  However, we focus on 
the specific question, how tight does the Bekenstein-Hawking relation 
constrain the form of the boundary Hamiltonian?  This question is 
more relevant to loop quantum gravity than string theory.

\section{Geometric particles}

We begin with a brief discussion of the statistical mechanics of a
non- interacting ensemble of distinguishable particles.  The
assumption of distinguishability makes this system a bit unusual in
that garden variety systems in statistical mechanics satisfy either
Bose-Einstein or Fermi-Dirac statistics.  This subsection also serves
to fix notation.  The results of this analysis will then be applied to
our geometric particles in Section \ref{ent}.

\subsection{Distinguishable particles and low temperatures} 
\label{cond}

Consider a system of distinguishable particles.  With each state of
definite energy, $\nu$, we associate a probability proportional to the
Boltzmann factor
\begin{equation} 
P_{\nu} \sim e^{-\beta(E_{\nu} - \mu N_{\nu})}, 
\end{equation} 
where $E_\nu$ is the total energy and $N_{\nu}$ is the particle number
of the state.  This assumption amounts to modeling the system as a
grand canonical ensemble.  We assume two characteristics of the
Hamiltonian.  First, the particle interactions are negligible. 
Second, the Hamiltonian has a discrete set of energy levels with
increasing energies $\{\epsilon_1, \epsilon_2, \dots\}$.  These levels
may be degenerate, so we denote the respective (finite) degeneracies
$\{g_1, g_2, \dots\}$.  The total Hamiltonian for the system cleanly
splits into a sum over the individual energies of the particles. 
Notice that we do not assume a precise form for the energy of the
particles.

A state of the system $\nu$ is uniquely specified (up to energy-level
degeneracy) by the number of particles $N_{\nu}$ and an ordered
$N_{\nu}$-tuple of integers $\{n_i\}$, where $n_i$ is the state label
of the $i$th particle (for $i = 1,2,\dots,N_{\nu}$).  Thus, $E_{\nu} =
\sum_{i=1}^{N_{\nu}} \epsilon_{n_i}$.  It is a simple matter to compute the
partition function
\begin{equation} 
\label{eq:gcz} 
\begin{split} 
{\cal Z} &= \sum_{ \{ \nu \} } e^{-\beta(E_{\nu} - \mu N_{\nu})} 
= \sum_{N=1}^{\infty} e^{\beta \mu N} \sum_{ \{ n_{i} \} } 
\prod_{i=1}^{N} g_{n_i} e^{-\beta \epsilon_{n_i}} 
\\ 
&= \frac{fz}{1-fz}, 
\end{split} 
\end{equation} 
where 
\begin{equation} 
f = \sum_{n=1}^{\infty} g_n e^{-\beta \epsilon_{n}} 
\end{equation} 
is the single particle partition function and $z=e^{\beta \mu}$ is the
fugacity.  With ${\cal Z}$, we can compute expectation values for the
total particle number, $N$, and the occupation number for the $k$th
energy level, $N_k$
\begin{eqnarray} 
\vev{N} &=& \frac{1}{\beta} \deriv{\ln {\cal Z}}{\mu} = \frac{1}{1 - fz} 
\label{eq:gcn} \\ 
\vev{N_{k}} &=& - \frac{1}{\beta} \deriv{\ln {\cal Z}}{\epsilon_{k}} = 
\frac {g_k e^{-\beta \epsilon_{k}}}{f} \left( \frac{1}{1-fz} \right). 
\label{eq:gcnk} 
\end{eqnarray} 
Therefore the fraction of particles in the $k$th state is strictly 
a function of the temperature of the system (independent of $\mu$) 
\begin{equation} 
\frac{\vev{N_{k}} }{\vev{N}} = \frac{g_k e^{-\beta \epsilon_{k}}}{f}. 
\end{equation}
Since 
\begin{equation} 
\label{eq:ratio1} 
\frac{\vev{N_{2}}}{\vev{N_{1}}} = \frac{g_{2}}{g_{1}} e^{-\beta (\epsilon_
{2} - \epsilon_{1})}, 
\end{equation} 
it is easy to see that for sufficiently low temperatures (as
determined by the natural scale of the theory), the system condenses
into the ground state.  This is the expected result; turning down the
temperature causes particles to fall into their lowest energy states. 
What may be surprising is that there exists a theory for which all
observed temperatures are in this regime.  That theory is quantum
gravity and we will see in the next section how this drives the
proportionality between area and entropy.

The reader might wonder why we see any interesting behavior at all for
such low temperatures, since the particles are in their ground states. 
However, since our system is a grand canonical ensemble, rather than
merely canonical ensemble, the system may still exhibit interesting
behavior in the form of particle number fluctuations (fluctuations in
$N$).  This may be derived with the relative dispersion relation
\begin{equation} 
\frac{\vev{\Delta N^{2}}}{\vev{N}^{2}} = 1 - 
\frac{1}{\vev{N}} 
\end{equation} 
Although the system condenses into the ground state fluctuations in
particle number remain significant.  At sufficiently low temperatures,
the relative dispersion goes as $1 - 1/\vev{N_{1}}$ so that particle
fluctuation is due to fluctuations in the number of particles in the
ground state.

\subsection{Entropy and Area} 
\label{ent}

We now show how the statistical mechanics of geometric particles leads
to the proportionality of entropy and area.  To accomplish this we must
characterize states of the system as observed from the exterior. 
Following \cite{carlobh}, given a foliation of spacetime the horizon is a
closed two dimensional surface.  On account of the horizon, different
states of the interior are regarded as physically indistinguishable
from the outside.  Thus, the quantum state of the system is specified
entirely by the state of the boundary \cite{carlobh,abk}.  Further, we
consider distinguishable particles since local geometric
fluctuations on the horizon relative to the exterior geometry are
diffeomorphism invariant \cite{carlobh}.

The natural scale of the theory is set by the Planck energy, $E_{\rm
P} \sim 10^{27} {\rm eV}$, so that the natural temperature scale of the
theory is the Planck temperature $T_{\rm P} \sim 10^{32} {\rm K}$.  We
previously found that condensation occurs when temperatures are
``sufficiently low.''  For quantum gravity this is $T \ll T_{P}$, a
condition that is certainly satisfied for all observed temperatures. 
Since the spacing between the ground and excited state energy levels
is on the order of the Planck energy, Eq.  (\ref{eq:ratio1}) tells us
that
\begin{equation} 
\frac{\vev{N_{2}} }{\vev{N_1}} \approx e^{-T_{\rm P}/T}. 
\label{eq:ratio} 
\end{equation} 
Temperatures commonly found in the universe are miniscule in
comparison to the Planck temperature, so the ratio in Eq. 
(\ref{eq:ratio}) is, for all intents and purposes,
zero.\footnote{Since $\epsilon_k > \epsilon_2$ for $k>2$, it follows
that $\vev{N_k}/\vev{N_1} = 0$ as well.} Thus, we come to the key
result that all of the particles condense into the ground state; at
the quantum gravity scale, the universe is such a chilly place that
the geometry of space ``freezes" into the lowest energy level.

It is straightforward to compute the quantities needed to check the
entropy- area relation.  For low temperatures, each particle rests in
the ground state, so the average energy simply scales with $\vev{N}$
\begin{equation}
\vev{E} = \epsilon_1 \vev{N}, \quad \mbox{for $T \ll T_{\rm P}$}.
\end{equation}

Now let's take a careful look at the area of the surface.  Since these
are non-interacting, geometric particles, each contributes
independently to the area.  Thus, for a state $\nu$ of the system
$A_{\nu} = \sum_{i=1}^{N_{\nu}} a_i$, where $a_i$ is the area
contribution of the $i$th particle.  Upon condensation, each particle
will be found with equal likelihood in one of the $g_1$ degenerate
ground states.  Note that each of these states could possibly yield a
different area contribution (although, they certainly do not have to),
so there could be anywhere from 1 to $g_1$ ``area levels" within the
ground state energy level.  Let $\vev{a}_{1}$ be the mean area
contribution among these $g_1$ states.  Clearly, $\vev{a}_{1}$ will be
a constant of the system, independent of temperature or particle
number.  We find that
\begin{equation}
\vev{A} = \vev{a}_{1} \vev{N}, \quad \mbox{for $T \ll T_{\rm P}$},
\end{equation}
so the area, too, scales with $\vev{N}$ at low temperatures.  Our
regime is quantum gravitational, so the constant of proportionality,
$\vev{a}_{1}$, is approximately equal to the Planck area $l^{2}_{P} \sim
10^{-70} {\rm m}^2$.  The immediate consequence is that all systems
with bounding area large compared with the Planck area -- what we call
``macroscopic" -- satisfy $\vev{N} \gg 1$.

We now have the following exact result for the entropy 
\begin{equation} 
\label{eq:entropy} 
\begin{split} 
S &= - k \sum_{\nu} P_{\nu} \ln P_{\nu} \\ 
&=k \left[ \beta \vev{E} - \mu \vev{N} + \ln (\vev{N} - 1) \right]. 
\end{split} 
\end{equation} 
For temperatures much lower than the Planck temperature, this becomes 
\begin{equation} 
S \simeq k \ln g_1 \vev{N} + k \ln \vev{N} - k \ln 
\left(1 - \frac{1}{\vev{N}} \right) \left( \vev{N} - 1 \right). 
\end{equation} 
Or, rewriting this in terms of the expectation value of the area we 
have 
\[ 
S \simeq \frac{k \ln g_1}{\vev{a}_{1}} \vev{A} + k \ln
\frac{\vev{A}}{\vev{a}_{1}} - k \ln \left (1 -
\frac{\vev{a}_{1}}{\vev{A}} \right) \left( \frac{\vev{A}}{\vev{a}_{1}}
- 1 \right).
\] 
It is clear what happens for macroscopic bounding surfaces ($\vev{A} 
\gg \vev{a}_{1}$) 
\begin{equation} 
\label{eq:entarea} 
S \simeq k \frac{\ln g_1}{\vev{a}_{1}} \vev{A} + k \ln
\frac{\vev{A}}{\vev{a}_{1}}.
\end{equation} 
The proportionality between entropy and area is recovered!  The first
correction term is logarithmic.  Note that the relation holds,
irrespective of the form of the microscopic dynamics.

\section{Conclusions} 
\label{sum}

In summary, we have shown that for temperatures $T \ll T_{P}$ and for
macroscopic black holes ($A \gg l^{2}_{P}$), a gas of
geometric particles will exhibit the entropy-area proportionality
(with logarithmic correction) irregardless of the details of the
dynamics.  This occurs when (i) the particles are distinguishable (ii)
the particles are non-interacting.  

We reach this result from the following steps:
\begin{itemize}
\item The quantum state of the system is specified entirely by the 
state of the boundary theory. 
\item With each state of definite energy, 
$\nu$, we associate the ``grand canonical" probability 
\begin{equation} 
P_{\nu} \sim e^{-\beta(E_{\nu} - \mu N_{\nu})}. 
\end{equation}
\item Each particle is noninteracting and may occupy a number of
discrete, possibly degenerate, energy levels with increasing energies. 
\item The difference in energy between the ground state and the first
excited state is on the order of the Planck energy; that is,
$\epsilon_2 - \epsilon_1 \approx E_{\rm P}$.  
\item The average area
of the surface, $\vev{A}$, is proportional to the average particle
number, $\vev{N}$.  Since our regime is quantum gravitational, the
constant of proportionality is approximately equal to the Planck area. 
Thus, any macroscopic system requires $\vev{N} \gg 1$.
\end{itemize} 
The entropy-area result follows. Consider, now, a few things we are 
{\em not} assuming: 
\begin{itemize} 
\item We do not assume any particular form of the area and energy
operators.  The result is insensitive to the microstructure of the
theory.  \item Nor do we assume any relationship between the energy
and area contributions of a single particle.\footnote{This is unlike
Ref.  \cite{gravstatmech} in which a precise relation {\em was}
assumed.} Indeed, a degenerate energy level could even correspond to
many different areas.
\end{itemize}

In regards to previous work, we note that Eq.  (\ref{eq:entarea}) does
not explicitly mention the Immirzi parameter $\gamma$ \cite{I}.  In fact, it
has merely been absorbed into the constant $\vev{a}_{1}$.  If we use
the appropriate values for the degeneracy and area coming from loop
quantum gravity, we find that (leaving out the logarithmic correction term)
\begin{equation} 
S \simeq k \frac{\ln 2}{4 \pi \sqrt{3} \gamma l_{P}^{2}} \vev{A}, 
\end{equation} 
which is precisely what was found in the isolated horizons black hole
calculation \cite{abk}.  (We use $l_{P} = \sqrt{\hbar G}$.) This is
despite the fact that the calculations are based on an entirely
different ensembles and statistical weights (although see below).

An interesting consequence of the condensation of particles into the
lowest energy level is that our grand canonical probability factor
becomes
\begin{equation}
e^{-\beta(E - \mu N)} \simeq e^{-\beta(\epsilon_1 - \mu) N} =
e^{-\alpha A}
\end{equation}
where we used the low temperature condition in the first equality. 
So, in the limit $T \ll T_{P}$ we recover the
statistical weight used in the ``area canonical ensemble" of Ref. 
\cite{kirill1}.  Further,
\begin{equation}
\alpha = \frac{\beta(\epsilon_1 - \mu)}{\vev{a}_{1}} \simeq \ln
g_{1}/\vev{a}_{1}.
\end{equation}
Thus, for macroscopic configurations the weight reduces to $g_{1} \exp
(-A/\vev{a}_{1})$.  

The argument presented in this paper indicates that, although full
quantum results must match semiclassical ones, using the entropy-area
relation offers little guidance to the form of the microdynamics.  There
is a huge universality class of theories which reproduce the
proportionality of entropy and area.  Thus, there are many
Hamiltonians which meet the very mild test of matching the
entropy-area proportionality.  Second, turning the argument around,
since the $S \propto A$ is true for a wide class of theories in the
low temperature and large area (compared to the Planck scale)
regime, a theory which merely matches this relation may not offer a
glimpse into Planck scale physics.

\end{document}